\documentstyle[12pt]{article}
\global\parskip 6pt

\normalsize

\let\oldtheequation=\theequation
\def\doteqs#1{\setcounter{equation}{0}
            \def\theequation{{#1}.\oldtheequation}}

\newcounter{sxn}
\def\sx#1{\addtocounter{sxn}{1} \bigskip\medskip \goodbreak
\noindent{\large\bf
\centerline{\thesxn.~~#1}} \nobreak \medskip}
\def\sxn#1{\sx{#1} \doteqs{\thesxn}}

\newcounter{axn}

\def\br{\begin{thebibliography}{abc}}
\def\er{\end{thebibliography}}

\def\be{\begin{equation}}
\def\ee{\end{equation}}
\def\bea{\begin{eqnarray}}
\def\eea{\end{eqnarray}}

\begin{document}
\begin{flushright}
\hfill{SINP-TNP/01-18}\\
%\hfill{hep-th/0102051}
\end{flushright}
\vspace*{1cm}
\thispagestyle{empty}
\begin{center}
{\large\bf Bound States in One-Dimensional Quantum
$N$-Body\\ Systems  with Inverse Square Interaction}\\
\end{center}
\bigskip
\begin{center}
B. Basu-Mallick\footnote{Email: biru@theory.saha.ernet.in}
and Kumar S. Gupta\footnote{Email: gupta@theory.saha.ernet.in}\\
\vspace*{.75cm}
{\em Theory Division\\
Saha Institute of Nuclear Physics\\
1/AF Bidhannagar, Calcutta - 700064, India}\\
\end{center}
\vskip.5cm

\begin{abstract}

We investigate the existence of
bound states in a one-dimensional quantum system of $N$ identical
particles interacting with each other through an inverse square potential. 
This system is equivalent to the Calogero model without the confining term. 
The effective Hamiltonian of this system in
the radial direction admits a one-parameter family of self-adjoint
extensions and the negative energy bound states occur when most general  
boundary conditions are considered. We find that these bound states exist 
only when $N=3,4$ and for certain values of the system parameters. 
The effective Hamiltonian for the system is related to the Virasoro algebra
and the bound state wavefunctions exhibit a 
scaling behaviour in the limit of small inter-particle separation.

\end{abstract}
\vspace*{1cm}
\begin{center}
September 2001
\end{center}
\newpage

\baselineskip 17pt

\sxn{Introduction} 

One dimensional many body quantum systems where the  2-body potential falls
as the inverse square of the inter-particle distance 
exhibit rich mathematical structures and often lead to interesting
physical applications [1-6]. This problem is equivalent to the 
Calogero model without the confining term. The spectrum of such
a system is normally described in terms of scattering states alone
\cite{calo3}.

It is however known that 
2-body systems on a half line containing an interaction term of the type
$g \over x^2$, where $g$ is the 2-body coupling and $x$
is the inter-particle distance, admit negative energy bound states 
for a certain range of parameter $g$ \cite{meetz,narn}. 
These bound states owe
their existence to self-adjoint extensions \cite{reed} of the corresponding
Hamiltonian and have recently found physical applications \cite{trg,ksg1}.

Another feature of these 2-body systems is that their corresponding 
Hamiltonians are related to the Virasoro algebra \cite{kac}. Although the
Hamiltonians themselves are not elements of the Virasoro algebra they do
belong to the corresponding enveloping algebra. They are thus contained in 
the representation space of the Virasoro algebra. This fact has 
interesting implications for physics of black holes \cite{ksg1}.

It is therefore natural to ask if these properties of the 2-body systems can
be generalized to the $N$-body case. In this Letter we provide a detailed 
analysis of this problem. We consider a system of $N$ identical particles of
mass $m$ interacting with each other through a potential of the type 
$g \over x^2$ where $x$ is the interparticle distance.
The 2-body coupling $g$ is assumed to be the same for all pairs of
particles. 
The effective potential describing this system is found to be 
a function of $g$ and $N$. 
The corresponding effective Hamiltonian  
is shown to admit self-adjoint extensions labelled by $e^{i z}$ where
$z \in R$.
Different values of the parameter $z$ leads to inequivalent quantizations of
the system. The spectrum of the system is in general a function of $g$,
$N$, and $z$. We show that for certain values of these parameters the system
admits negative energy bound states. We also point out that the 
effective Hamiltonian for the $N$-body system 
is contained in the enveloping algebra of the Virasoro algebra and
thus have a well defined action on the representation space of the Virasoro 
algebra. The wave functions describing the negative energy bound states
exhibit scaling behaviour in a certain regime.

The organization of this paper is as follows. In Section 2, we study the 
$N$-body system where the pairwise potential between 
any two particles is of the
inverse square type and obtain the effective Hamiltonian for the system.
In Section 3 we quantize the effective Hamiltonian and find the
corresponding self-adjoint extensions. The negative energy bound states are
obtained for certain values of the parameters of the problem.
In Section 4 we indicate the connection of this system with Virasoro algebra
and discuss the scaling behaviour of the bound states.
We conclude in Section 5 with a summary and a outlook.

%\newpage
\sxn{Effective Hamiltonian for the $N$-body System}

	Let us consider a system of $N$ identical particles of mass $m$
in one dimension whose Hamiltonian is given by
\be
H = - \sum^{N}_{i=1} \frac{{\partial}^2}{\partial x_i^2} +  \sum^{N}_{i=2}
\sum^{i-1}_{j=1} \frac{g}{(x_i - x_j)^2},
\ee
where $x_i$ is the coordinate of the $i^{\rm th}$ particle.
The interaction strength $g$ between any two particles 
is assumed to be the same for all pairs of particles and units have been 
chosen such that $2 m {\hbar}^{- 2} = 1$.
If $g < - \frac{1}{2}$
the two-body system is unstable and the corresponding ground state energy is 
unbounded from below \cite{land}. In what follows we shall assume that 
$g \ge - \frac{1}{2}$.

	We are interested in finding normalizable solutions of the
eigenvalue problem 
\be
H \psi = {\cal {E}} \psi
\ee
especially when ${\cal {E}} < 0$.
Following \cite{calo3}, we consider the above eigenvalue equation
in a sector of configuration
space corresponding to a definite ordering of particles given by
$x_1 \geq x_2 \geq
\cdots \geq x_N$. The translation-invariant
 eigenfunctions of the Hamiltonian $H$ can be written as  
\be
\psi = y^{a + \frac{1}{2}} \phi (r) P_k (x),
\ee
where
\bea
y &=& \prod^N_{i=2}\prod^{i-1}_{j=2} (x_i - x_j),\\
r^2 &=& \frac{1}{N} \sum^{N}_{i=2}\sum^{i-1}_{j=1} (x_i - x_j)^2,\\
a &=& \pm \frac{1}{2} ( 1 + 2 g)^{1 \over 2}
\eea
and $P_k (x)$ is a homogeneous  as well as translation-invariant 
polynomial of degree $k$ which satisfy the
equation
\be
\left[ \sum^{N}_{i=1}\frac{{\partial}^2}{\partial x_i^2} 
+ 2 (a + \frac{1}{2}) \sum^{N}_{i=2}\sum^{i-1}_{j=1} \frac{1}{(x_i -
x_j)}  \left( \frac{{\partial}}{\partial x_i} - 
\frac{{\partial}}{\partial x_j} \right) \right] P_k (x) = 0. 
\ee
From Eqn. (2.3) we see that the condition
$a + \frac{1}{2} \geq 0$ must be satisfied for the wavefunction 
to be well behaved when $x_i \rightarrow x_j$. However, for the case
$a + \frac{1}{2} = 0$,  (2.1) evidently reduces to a free Hamiltonian of $N$ 
particles. At present we shall consider only the case
  $a + \frac{1}{2} > 0$ for which the wavefunction (2.3)
vanishes at the limit $x_i \rightarrow x_j$ \cite{poly}.

Substituting Eqn. (2.3) in Eqn. (2.2) and using Eqn. (2.7) we get
\be
- \frac{d^2 \phi}{dr^2} - (1 + 2 \nu )
\frac{1}{r} \frac{d \phi}{d r} = {\cal {E}} \phi,
\ee 
where 
\be
\nu = k + \frac{1}{2}(N - 3) + \frac{1}{2} N (N-1)(a + \frac{1}{2}).
\ee
We note that for $N \geq 3$, $ \nu $ is a positive definite quantity.
Let us define a function $\chi (r)$ through the transformation
\be
 \phi (r) = 
{\rm {exp}} \left[ - \frac{1}{2} \int_1^r (1 + 2 \nu )
\frac{1}{r}  dr \right] \chi (r) = r^{- (\frac{1}{2} + \nu )} \chi (r). 
\ee
Substituting Eqn. (2.10) in Eqn. (2.8), we get
\be
\left[ -\frac{d^2 }{dr^2} + \frac{\nu^2 - \frac{1}{4}}{r^2} \right] \chi (r)
= {\cal {E}} \chi (r).
\ee
As mentioned earlier, we are interested in solutions of Eqn. (2.2) where
${\cal {E}} < 0$. Writing ${\cal {E}} = - E$  where $E > 0$, we can rewrite
Eqn. (2.11) as 
\be
\tilde{H} \chi(r) = - E \chi(r)
\ee
where
\be
\tilde{H} = -\frac{d^2 }{dr^2} + \frac{\tilde{g}}{r^2}
\ee
and $\tilde{g} = \nu^2 - \frac{1}{4}$. 
$\tilde{H}$ defines the effective Hamiltonian for the system and $\tilde{g}$
is the effective coupling.

It may be noted that,
the variable $ r $ in Eqn. (2.5) 
takes values in the positive real axis 
$ R^+ \equiv [0, \infty ]$. In fact,  $r$  represents the 
radial variable associated with $N-1$ number of 
translation-invariant `Cartesian'
coordinates (i.e. Jacobi coordinates), 
which are obtained from the $x_i$'s
through the elimination of center-of-mass coordinate \cite {calo3}.
 Consequently, the operator $\tilde{H}$ in Eqn. (2.13)
defines the effective Hamiltonian in the radial direction for the $N$-body 
system under consideration. 
By using Eqns. (2.3) and (2.10), the wavefunction of the $N$-body system can
be expressed in terms of the eigenfunction of the effective Hamiltonian as 
\be
\psi = y^{a + \frac{1}{2}} r^{-( \frac{1}{2} + \nu )}  P_k (x) \chi(r).
\ee
 
We would next like to find the measure for which the eigenfunctions $\chi
(r)$ of the effective Hamiltonian $\tilde{H}$ would be square integrable.
Let the angular variables 
 associated with $N-1$ translation-invariant Cartesian 
coordinates be denoted by $\Omega_i$ ($i\in [1,N-2]$) \cite{calo3}. 
The 
relative distance between any two particles can be expressed  in the form
 $x_i -x_j = r X_{ij}(\Omega)$,
where  $X_{ij}(\Omega)$ is some function of these angular variables 
alone. Substituting this form of $x_i -x_j$ into Eqn. (2.14) and also using 
the fact that $P_k (x)$ is a translation-invariant homogeneous  
polynomial of degree $k$, $\psi $ can be factorized as 
$\psi = \xi (r) F(\Omega)$ where
 
\be
\xi (r) =  r^{ {N(N-1)\over 2} (a+\frac{1}{2}) + k  -( \frac{1}{2} + \nu ) } 
  \chi(r) 
\ee
and $F(\Omega)$ depends on the angular variables alone.
Since $\psi $ is a normalizable eigenfunction depending on 
 $N-1$ number of translation-invariant Cartesian coordinates, the integral 
$\int_0^\infty \xi^*(r) \xi (r) r^{N-2} dr $ must be finite and
$\xi (r) \in L^2[R^+, r^{N-2} dr]$.
Now, using Eqns.(2.15) and (2.9) we get,
\be
\int_0^\infty \xi^*(r) \xi(r) r^{N-2} dr = 
%\int_0^\infty \chi^*(r) \chi(r) 
%  r^{ N(N-1) (a+\frac{1}{2}) + N-2 +2k  -( 1 +2 \nu ) } dr = 
\int_0^\infty \chi^*(r) \chi(r) dr.
\ee
Thus $ \int_0^\infty \chi^*(r) \chi(r) dr $ is also finite and 
$\chi (r) \in L^2[R^+, dr]$.
In the next
Section we shall study the eigenvalue problem for the effective Hamiltonian.

\sxn{Self-adjoint Extension and Bound States}

	The operator $\tilde{H}$ of Eqn. (2.13) 
belongs to a general class of objects
known as  unbounded linear differential operators on a Hilbert space
\cite{reed}. We start by summarizing some properties of these operators
which are needed for our purpose.

	Let $T$ be an unbounded differential operator acting on a Hilbert
space ${\cal H}$ and let $(\gamma , \delta )$ denote the inner product 
of the elements $\gamma , \delta \in {\cal H}$.
By the Hellinger-Toeplitz theorem \cite{reed}, $T$  has a well defined 
action only on a dense subset $D(T)$ of the Hilbert space  ${\cal H}$. 
$D(T)$ is known as the domain of the operator $T$. Let $D(T^*)$ be the set
of $\phi \in {\cal H}$ for which there is an $\eta \in {\cal H}$ with
$(T \xi , \phi) = (\xi , \eta )~ \forall~ \xi \in D(T)$. For each such
$\phi \in D(T^*)$ we define $T^* \phi = \eta$. $T^*$ is called the adjoint
of the operator $T$ and $D(T^*)$ is the corresponding domain of the adjoint.

The operator $T$ is called symmetric or Hermitian if $T \subset T^*$,
i.e. if $D(T) \subset D(T^*)$ and $T \phi = T^* \phi~ \forall~ \phi \in 
D(T)$. Equivalently, $T$ is symmetric iff $(T \phi, \eta) = (\phi, T \eta)
~ \forall ~ \phi, \eta \in D(T)$. The operator $T$ is called self-adjoint 
iff $T = T^*$ and $D(T) = D(T^*)$.

We now state the criterion to determine if a symmetric operator $T$ is 
self-adjoint. For this purpose let us define the deficiency subspaces 
$K_{\pm} \equiv {\rm Ker}(i \mp T^*)$ and the 
deficiency indices $n_{\pm}(T) \equiv
{\rm dim} [K_{\pm}]$. $T$ is (essentially) self-adjoint iff 
$( n_+ , n_- ) = (0,0)$.
$T$ has self-adjoint extensions iff $n_+ = n_-$. There is a one-to-one
correspondence between self-adjoint extensions of $T$ and unitary maps
from $K_+$ into $K_-$. Finally if $n_+ \neq n_-$, then $T$ has no
self-adjoint extensions.

We now return to the discussion of the effective Hamiltonian $\tilde{H}$. 
This is an unbounded differential operator defined in 
$R^+ $. $\tilde{H}$ is a symmetric operator on the domain 
$D(\tilde{H}) \equiv \{\phi (0) = \phi^{\prime} (0) = 0,~
\phi,~ \phi^{\prime}~  {\rm absolutely~ continuous} \} $.
It is known
that for $\tilde{g} \geq \frac{3}{4}$, 
$\tilde{H}$ has deficiency indices (0,0) and is 
(essentially) self-adjoint on the domain $D(\tilde{H})$.
For $ - \frac{1}{4} \leq \tilde{g} < \frac{3}{4}$, 
$\tilde{H}$ has deficiency
indices $ (1,1)$ \cite{hut}. In the latter case 
$\tilde{H}$ is not  self-adjoint on the 
domain $D(\tilde{H})$ but admits self-adjoint extensions.
The deficiency
subspaces  $K_{\pm}$ in this case are one dimensional and are spanned by
\bea
\phi_+ (r) &=& r^{\frac{1}{2}}H^{(1)}_\nu (re^{i \frac{ \pi}{4}}),\\ 
\phi_- (r) &=& r^{\frac{1}{2}}H^{(2)}_\nu (re^{-i \frac{ \pi}{4}})
\eea
respectively, where $H_\nu$'s are Hankel functions.
The unitary maps from $K_+$ into $K_-$ are parametrized by $e^{i z}$ where 
$z \in R$. The operator $\tilde{H}$ is self-adjoint in the domain
$D_z(\tilde{H})$ which contains all the elements of $D(\tilde{H})$ together
with elements of the form 
$\phi_+ (r) + e^{i z} \phi_- (r) $. Different values of the
parameter $z$ lead to inequivalent quantizations of the operator
$\tilde{H}$. 

	As mentioned earlier, we are interested in the negative energy
normalizable solutions of Eqn. (2.12).
For $\tilde{g} \geq \frac{3}{4}$ there are no such bound states \cite{narn}.
For each value of 
$\tilde{g}$ such that $ - \frac{1}{4} < \tilde{g} < \frac{3}{4}$, i.e.
$0 < \nu < 1$, the solution of Eqn. (2.12) is given by
\be
\chi (r) = B (\sqrt{E} r)^{\frac{1}{2}}[J_\nu (i \sqrt{E} r) - e^{i \pi \nu}
J_{- \nu} (i \sqrt{E} r)],
\ee
where $J_{\pm \nu}$ are Bessel functions and
$B$ is the normalization constant \cite{narn}.
In order to find the energy $E$, we use the fact that
if $\tilde{H}$ has to be self-adjoint, the eigenfunction 
$\chi (r)$ must belong to the domain $D_z(\tilde{H})$. 
Note that in the limit $r \rightarrow 0$,
\be
\phi_{+}(r) + e^{i z} \phi_{-}(r) \rightarrow \frac{i}{{\rm sin}\nu \pi}
\left [ \frac{r^{\nu + \frac{1}{2}}}{2^\nu} \frac{( e^{ - i  
\frac{3 \nu \pi}{4}} - e^{i (z +  \frac{3 \nu \pi}{4})})}{\Gamma (1 + \nu)}
+ \frac{r^{- \nu + \frac{1}{2}}}{2^{- \nu}}
\frac{( e^{ i
( z +  \frac{\nu \pi}{4} ) } - e^{-i \frac{ \nu \pi}{4}}) }{\Gamma (1 -
\nu)}
\right ]
\ee  
and
\be
\chi (r) \rightarrow B \left [ \frac{r^{\nu + \frac{1}{2}}}{2^\nu}
\frac{(\sqrt{E})^{\nu + \frac{1}{2}} e^{i  \frac{ \nu \pi}{2}}}{\Gamma (1 +
\nu)} - \frac{r^{- \nu + \frac{1}{2}}}{2^{- \nu}}
\frac{(\sqrt{E})^{- \nu + \frac{1}{2}} e^{i  \frac{ \nu \pi}{2}}}{\Gamma (1
- \nu)} 
\right ].
\ee
Thus if $\chi (r) \in D_z(\tilde{H})$, then the coefficients of 
of $r^{\nu + \frac{1}{2}}$ and $r^{- \nu + \frac{1}{2}}$
 in Eqns. (3.4) and
(3.5) must match. Comparing these coefficients we get
\be
E = \left [ \frac{{\rm sin}(\frac{z}{2} + 3 \pi \frac{\nu}{4})}
{{\rm sin}(\frac{z}{2} + \pi \frac{\nu}{4})} \right ]^{\frac{1}{\nu}},
\ee
Thus we see that for every value of $\tilde{g}$ within the range
$ - \frac{1}{4} < \tilde{g} < \frac{3}{4}$, $\tilde{H}$ admits a single 
bound state with wavefunction given by Eqns. (3.3) and bound state energy
given by ${\cal E} = - E$.

We shall now show that the existence of the negative energy bound states
for the original $N$-body system depends on $N$ and the degree $k$ of
the polynomial $P_k (x)$. To this end, let us recall that
$\nu $ must lie in the range $0 < \nu < 1$ for the negative energy bound
states to exist. This together with
Eqn. (2.9) gives
\be
0 <  
k + \frac{1}{2}(N - 3) + \frac{1}{2} N (N-1)(a + \frac{1}{2}) <  1.
\ee
Using Eqn. (3.7) along with the constraint $a + \frac{1}{2} > 0$
we get
\be
0 < a + \frac{1}{2} < \frac{5 - N - 2 k}{N (N - 1)}.
\ee
Let us consider the situation when $k = 0$. This corresponds to a class of 
eigenstates which includes the ground state
of the system \cite{calo2,calo3} and for $N=3$ corresponds to the case of
the zero angular momentum sector \cite{calo1,calo2}. In this case we have
\be
0 < a + \frac{1}{2} < \frac{5 - N}{N (N - 1)}.
\ee
Thus the 
allowed values of $N$ which satisfy the above constraint are $N = 3$ and 4.
These are the only cases where a negative energy bound state can appear in the
system. For $N=3$, the allowed ranges for $a$ and $g$ are given by 
$- \frac{1}{2} < a < - \frac{1}{6}$ and 
$ 0 > g > - \frac{4}{9} $ respectively. Similarly for $N=4$,
$- \frac{1}{2} < a < - \frac{5}{12}$ and $ 0 > g > - \frac{11}{72}$. 
From the above discussion it follows that negative energy bound states may 
occur only when the 2-body coupling $g$ in Eqn. (2.1) is attractive.

Let us now consider the case when $k > 0$. From Eqn. (3.8) it follows that
$5 - N - 2 k > 0$. There is no value of $N > 2$ which satisfies this   
constraint. Thus there are no negative energy bound states for $k > 0$.

We therefore see that the original $N$-body problem of
Eqn. (2.2) admits negative energy bound states when $k = 0$ and when
$N = 3$ or 4 with corresponding range of the coupling $g$ as described
above. Note that for $k = 0$, $P_k (x) = 1$. The full solution of Eqn. (2.2)
in this case is given by
\be 
\psi = B y^{a + \frac{1}{2}} r^{-( \frac{1}{2} + \nu )}
(\sqrt{E} r)^{\frac{1}{2}}(J_\nu (i \sqrt{E} r) - e^{i \pi \nu}
J_{- \nu} (i \sqrt{E} r)).
\ee
The corresponding energy has the value 
${\cal E} = - E$ where $E$ is given by Eqn. (3.6).

We end this Section with the following observations :

\noindent
1) In the discussion above we have seen that the constraint
$a + \frac{1}{2} > 0$ puts strong restrictions on the allowed values of $N$.
This constraint is needed to keep the wavefunction in Eqn. (2.3) well behaved
when $x_i \rightarrow x_j$. 
It is possible to consider a situation where there is a   
natural cutoff in the system which prevents any two particles coordinates 
from coinciding with each other.
In the presence of a natural cutoff, the wavefunction in Eqn. (2.3) would be
well behaved even if
$a + \frac{1}{2} < 0$. In such cases it seems likely that negative energy
bound states would exist for all values of $N$. 

\noindent
2)  
When $\tilde{g} = - \frac{1}{4}$, 
or equivalently $\nu =0$, 
Eqn. (2.12) admits an infinite number of bound states \cite{meetz,narn}.
In this case $a$ satisfies the condition
$a + \frac{1}{2} =  \frac{3 - N -2k}{N(N-1)}$, which together with the
constraint $a + \frac{1}{2} > 0$  
precludes any value of $N > 2$. However, when $N=3$ 
and $k=0$, the above condition yields $a = - \frac{1}{2}$ 
which corresponds to a system of free
particles with $g = 0$. 
Eliminating the centre-of-mass degree of freedom when $N=3$ and $g=0$, 
and using Jacobi coordinates $y_1$,
$y_2$ \cite{calo1}, Eqn.(2.2) can be expressed as 
\be 
- \left( 
 \frac{{\partial}^2}{\partial y_1^2} +  
 \frac{{\partial}^2}{\partial y_2^2}  \right) \psi =  {\cal {E}} \psi .
\ee
This represents the eigenvalue problem 
of two free particles in one dimension, or equivalently, 
one free particle in two dimension. Using polar coordinates
and restricting to the zero angular momentum sector, the radial part of Eqn.
(3.11) can be transformed in the form of Eqn. (2.12) with 
$\tilde{g} =  - \frac{1}{4}$. Following \cite{narn},
the solutions of Eqn.(3.11) can thus be obtained as 
\be 
\psi_n = K_0( \sqrt {- {\cal E}_n} r) , ~~~~ {\cal E}_n = 
- e^{ {\pi \over 2}(1-8n) 
\cot{z \over 2}} ,
\ee
where $r = \sqrt {y_1^2 + y_2^2} $, 
$ K_0( \sqrt {-{\cal E}_n} r) $ is the modified Bessel function 
 and $z$ represents the self-adjoint extension parameter. Thus we find that
the zero angular momentum sector of 
this system in Eqn. (3.11) admits an infinite number of 
negative energy bound states. 
This may seem surprising as the operator in Eqn.
(3.11) is normally assumed to be positive. This is however not true in the
presence of self-adjoint extensions which may break the positivity \cite{bonn}.

\noindent
3)Finally, when $\tilde{g} < - \frac{1}{4}$, the eigenvalue problem of Eqn.
(2.2) does not have a physically meaningful solution as the 
the ground state energy in this case is unbounded from below \cite{land}.
The effective Hamiltonian $\tilde{H}$ could still be made self-adjoint through
appropriate choice of a domain. This is however not sufficient for the
ground state energy to be well defined. The 2-body problem in this region of
the coupling has been analyzed using renormalization group techniques
\cite{ksg}. Investigation of the $N$-body problem in this case is still open.

%\newpage

\sxn{Connection to Virasoro Algebra}

The Hamiltonian of a 2-body system with inverse
square potential is known to be related to conformal symmetry and Virasoro
algebra
\cite{ksg1,fub}. The form of $\tilde{H}$ in Eqn. (2.13) suggests that the 
effective Hamiltonian for $N$-body system should also have similar 
property. Consider the Virasoro generators
$L_n = -x^{n + 1} \frac{d}{dx}$ and shift operators  
$P_m = \frac{1}{x^m}$ where $n,~m \in {\mathbf Z}$.
In terms of $L_{- 1}$ and $P_1$, $\tilde{H}$ can be factorized as  
\be
\tilde{H} = (- L_{-1} + \mu P_1) (L_{-1} + \mu P_1)
\ee
where $\mu = \frac{1}{2} + \nu$ \cite{ksg1}. Commutators of 
$\tilde{H}$ with $L_{- 1}$ and $P_1$ leads to other elements of the algebra
and continuing this process produces all the generators \cite{ksg1}.
In particular, the 
commutator of $\tilde{H}$ with a general element $L_{m}$ gives
\be
[L_m, \tilde{H}] = 2\mu(\mu-1)P_{2-m} - (m+1)(L_{-1}L_{m-1} + L_{m-1}L_{-1}).
\ee
The right hand side of Eqn. (4.2) contains products of the Virasoro
generators. While these 
product operators are not elements of the Virasoro algebra,
they do belong to the corresponding enveloping algebra. Thus $\tilde{H}$
is contained in the enveloping algebra of the Virasoro algebra and has a
well defined action on the associated representation space \cite{kac}.

The above discussion is valid for arbitrary values of
$\tilde{g}$. We now focus our
attention on the bound states described by Eqn. (3.3) which are obtained when
$ - \frac{1}{4} < \tilde{g} < \frac{3}{4}$. 
If the effective Hamiltonian $\tilde{H}$ is indeed
associated to a Virasoro algebra, it might be expected that the
wavefunctions exhibit a suitable scaling behaviour. The wavefunction
$\chi (r) $ in Eqn. (3.3) however has no clear scaling properties.
This is due to the fact that in the presence of the self-adjoint extension,
the scaling operator in general does not leave the domain of 
$\tilde{H}$ invariant \cite{trg}.
Consider now the behaviour of the wavefunction when all the particles in the
system are close to each other. In this limit of $r \rightarrow 0$, the
behaviour of the wavefunction $\chi (r) $ is given by Eqn. (3.5). From this
expression we see that for small values of $r$, the leading scaling
behaviour of the $\chi (r) $ is governed by $r^{-  \nu + \frac{1}{2}}$ 
and the wavefunction exhibits approximate scaling behaviour in this limit.

\sxn{Conclusion}

	In this paper we have investigated the existence of bound states in
 a one-dimensional 
$N$-body quantum system with inverse square interaction.
We have
shown that this system admits negative energy bound states when 
either
1) $N = 3$, $k=0$ and $ 0> g > - \frac{4}{9}$ or 2) $N = 4$, $k=0$ and
$ 0> g > - \frac{11}{72}$. Moreover, there exists only a single such state
for each value of allowed value of $g$. 
To our knowledge, this is the first demonstration 
of the existence of negative energy bound states in this system. 

These bound states have been obtained by analyzing the effective Hamiltonian
$\tilde{H}$. The existence of these
negative energy states is related to the self-adjoint extensions of 
$\tilde{H}$. Each choices of
the self-adjoint parameter $z$ leads to a different domain  on which
$\tilde{H}$ acts. Since the boundary condition associated with a given
physical
situation is encoded in the choice of the domain, the self-adjoint
parameter $z$ is directly related to the physics of the problem. This is
clearly seen by the fact that different values of $z$ lead to inequivalent
expressions for the bound state energy (cf. Eqn. (3.6)). We therefore
conclude that the negative energy bound states are found only when most 
general boundary conditions are considered. 

The effective Hamiltonian $\tilde{H}$ is contained in the 
enveloping algebra of the Virasoro algebra and has a well defined action on
the associated representation space.
This may seem to imply the existence of
scaling symmetry in the problem which should be reflected in the
properties of the bound states. The self-adjoint extensions of $\tilde{H}$
in general break the scaling symmetry. However, when all the particles
in the system are close to each other, the bound state 
wavefunction in Eqn. (3.5) exhibits an approximate scaling behaviour. 
This appears to be the only remnant of the scaling symmetry at the quantum
level.

It would be interesting to consider the system described here 
in the presence of a natural
cutoff which may lead to bound states for all values of $N$. 

\vskip 1cm
\noindent
{\bf Acknowledgements}

K.S.G. would like to thank A.P.Balachandran, Diptiman Sen and Siddhartha Sen
for discussions.

\newpage
\bibliographystyle{unsrt}

\end{document}